\documentclass[a4paper,fleqn] {cas-dc}

\usepackage[numbers]{natbib}
\usepackage{algorithm}
\usepackage{algpseudocode}

\begin{document}
\let\WriteBookmarks\relax
\def\floatpagepagefraction{1}
\def\textpagefraction{.001}
\shorttitle{Improving Floyd-Warshall APSP Algorithm}
\shortauthors{IH Toroslu}

\title [mode = title]{Improving The Floyd-Warshall All Pairs Shortest Paths Algorithm}                      

\author[1]{Ismail H. Toroslu}[orcid=0000-0002-4524-8232]


\address[1]{Dept. of Computer Eng., METU, Ankara, Turkey}

\begin{abstract}
%
%
The Floyd-Warshall algorithm is the most popular algorithm for determining the shortest paths between all pairs in a graph. It is very a simple and an elegant algorithm. However, if the graph does not contain any negative weighted edge, using Dijkstra's shortest path algorithm for every vertex as a source vertex to produce all pairs shortest paths of the graph works much better than the Floyd-Warshall algorithm for sparse graphs. Also, for the graphs with negative weighted edges, with no negative cycle, Johnson's algorithm still performs significantly better than the  Floyd-Warshall algorithm for sparse graphs. Johnson's algorithm transforms the graph into a non-negative one by using the Bellman-Ford algorithm, then, applies the Dijkstra's algorithm. Thus, in general the Floyd-Warshall algorithm becomes very inefficient especially for sparse graphs. In this paper, we show a simple improvement on the Floyd-Warshall algorithm that will increases its performance especially for the sparse graphs, so it can be used instead of more complicated alternatives.  
\end{abstract}




\maketitle

\section{Introduction}

Due to its simplicity, the Floyd-Warshall \cite{Floyd1962}-\cite{Warshall1962} algorithm is preferred over the Dijkstra's \cite{Dijkstra1959} (appying it for every vertex as a source) or Johnson's \cite{Johnson1977} algorithms for solving all pairs shortest paths problems most of the time. In general, for small sized problems, or, for dense graphs the Floyd-Warshall algorithm works better than its alternatives. However, since it has a rigid form, its complexity is always $O(N^3)$ (where $N$ is the number of vertices in the graph) regardless of the graph structure.  Thus, it is not flexible or adaptive for very large and sparse graphs, and for these kind of graphs, it cannot compete with its alternatives. In this paper, we propose a very simple modification to the Floyd-Warshall algorithm in order to make it suitable for large sparse graphs. 

The Floyd Warshall algorithm is based on a very smart recurrence relation. Let $w_{ij}$ to represent the initial weights on the entries of the adjacency matrix of the graph, and, $d_{ij}^k$ be the weight of the shortest path from vertex $i$ to $j$ using the vertices from the set  $\{1, 2, ..., k\}$ (thus, $d_{ij}^0$ corresponding to the $w_{ij}$ of the adjacency matrix), then, the following recurrence relation holds:

\begin{equation}
 d_{ij}^k =\left \{
 \begin{array} {ll}
 w_{ij} & \mbox{if } k=0 \\
 min(d_{ij}^{k-1}, d_{ik}^{k-1}+d_{kj}^{k-1}) & \mbox{if } k>0
 \end{array}
 \right.
\end{equation}

We will also use the notation $D^k$ to represent the matrix corresponding to the calculation of $d_{ij}^k$, i.e., the distances of shortest paths obtained by using the vertices $\{1, 2, ..., k\}$ as intermediate vertices.

Like other shortest path algorithms, the main operation of this recurrence relation corresponds to the following relaxation operation:\\

if $((d_{ik}+d_{kj})<d_{ij})$ then $d_{ij} \leftarrow d_{ik}+d_{kj}$ \\

The recurrence relation basically describes in which order this relaxation operation is going to be executed. According to this recurrence relation, at each iteration, vertices are added into a set (represented by a superscript $k$) one by one that can be used in the construction of the shortest paths.  Naturally, this process ends when all vertices are included. It is  assumed that $w_{ii}=0$, and there is no self edge, and also, if there is no edge between the vertices $i$ and $j$, then $w_{ij}=\infty$.
This recurrence relation is typically implemented as shown in the FW Algorithm.

\begin{algorithm}
  \begin{algorithmic}[1]
  \State $N$: number of vertices
  \State $A[1..N,1..N]$: adjacency matrix ($d_{ij}^0$)
\Procedure {FW} {$N, A[1..N,1..N]$}
  \For {$k{\bf :} 1 \to N$} 
  \For {$i{\bf :} 1 \to N$} 
  \For {$j{\bf :} 1 \to N$} 
    \If{$A[i,j]>A[i,k]+A[k,j]$} 
      \State {$A[i,j] \gets A[i,k]+A[k,j]$}
    \EndIf
    \EndFor
    \EndFor
  \EndFor
\EndProcedure
 \caption{FW}\label{FW}
\end{algorithmic}
\end{algorithm}

The outer loop of the Algorithm 1 (the Floyd-Warshall algorithm) corresponds to the iterations of the above recurrence relation. The same matrix ($A[*,*]$) is used for going from iteration $k-1$ to $k$. Thus, after each iteration of the outer loop, the entries of the matrix $A$ (such as $A[i,j]$) represents $d_{ij}^k$. The second and the third loops are used to determine the new distance values between all pairs of the vertices of the graph, for the current iteration $k$ of the recurrence relation. 

As it can be seen from this implementation, even for sparse graphs, i.e., graphs with vertices having very few outgoing and incoming edges, the second and the third loops will explore the whole matrix containing entries mostly with $\infty$ values. Since, the condition of the {\bf if} statement will not be satisfied when either   $A[i,k]$ or $A[k,j]$ is $\infty$, most of these explorations will be a useless. 
Thus, rather than exploring the whole matrix to discover very few useful entries, these loops should be modified such that only useful entries are explored. In other words, all of the explored entries of the matrix  should contain non $\infty$ values. 

In this paper we propose an improvement on this recurrence relation and on the corresponding algorithm.In the next section we introduce our improved algorithm. Then, we also show with some experiments its effectiveness, before the conclusion.

\section{Improved Floyd-Warshall Algorithm: Eliminating Useless Relaxation Attempts}

First, we define the useless relaxation attempt, and then, we show how all of them can easily be avoided.

\newdefinition{definition}{Definition}
\begin{definition}
{\bf Useless Relaxation Attempt:} If either $d_{ik}$ or $d_{kj}$, or both of them, have $\infty$ value (i.e. there is no edge between these vertices), then the attempt of the relaxation operation (if $((d_{ik}+d_{kj})<d_{ij})$ then $d_{ij} \leftarrow d_{ik}+d_{kj}$) is {\em useless}.
\end{definition}

It is very easy to find the {\em useful} entries of the matrix, by only exploring the outgoing and the incoming edges of the vertex ($k$) that is currently being processed. That means, the second loop should only explore the entries corresponding to the incoming edges of the vertex $k$, and the third loop should only explore the entries corresponding to the outgoing entries of the vertex $k$. Moreover, during the application of an iteration of the recurrence relation, new paths may be formed between pairs of vertices that did not have any path before that iteration. This requires, properly adding the edges corresponding to these new paths to the incoming and outgoing edge lists of the all the vertices being effected, in addition the converting an $\infty$ entry of the matrix to a non-$\infty$ value. As a result, in later iterations, the {\em useful} entries corresponding to these matrix entries will be explored since they will correspond to the incoming and outgoing edge lists of some vertices.

In order to prevent {\em useless relaxation attempts}, the Floyd-Warshall recurrence relation has to be modified such that, only the incoming and the outgoing edges of the vertex $k$ should be used at the iteration $k$. Let $in_\alpha^\beta$ ($out_\alpha^\beta$) represent the list of the vertices on the other end of the incoming (outgoing) edges of the vertex $\alpha$ in iteration $\beta$ (i.e., the matrix $D^\beta$). Then, we can rewrite Floyd-Warshall recurrence relation as follows:

\begin{equation}
 d_{ij}^k =\left \{
 \begin{array} {ll}
 w_{ij} & \hspace*{-5cm} \mbox{if } k=0 \\
 min(d_{ij}^{k-1}, d_{ik}^{k-1}+d_{kj}^{k-1}) & \\
 ~~~~~~~~~~~~~~~~~\mbox{if } k>0 \wedge i \in in_k^{k-1} \wedge j \in out_k^{k-1}
 \end{array}
 \right.
\end{equation}
\begin{equation}
 in_j^k = \left \{
 \begin{array}{ll}
 in_j^0 \bigcup \{i\} & \mbox{if } k= 0 \wedge w_{ij} \neq \infty \wedge i \neq j\\
 in_j^{k-1} \bigcup \{i\} & \mbox{if } k>0 \wedge d_{ij}^{k-1} = \infty \wedge d_{ij}^k \neq \infty \\
 \end{array}
 \right.
\end{equation}
\begin{equation}
 out_i^k = \left \{
 \begin{array}{ll}
 out_i^0 \bigcup \{j \} & \mbox{if } k= 0 \wedge w_{ij} \neq \infty \wedge i \neq j\\
 out_i^{k-1} \bigcup \{j\} & \mbox{if } k>0 \wedge d_{ij}^{k-1} = \infty \wedge d_{ij}^k \neq \infty \\
 \end{array}
 \right.
\end{equation}

\newtheorem{theorem}{Theorem}
\begin{theorem}
Recurrence relations corresponding to the equations (2-4) prevents all {\em useless relaxation attempts}.
\end{theorem}
\newproof{pf}{Proof}
\begin{pf}
Since in each step $k$, the relaxation operations are applied to $ij$ pairs only if there is an edge from vertex $i$ to the vertex $k$ ($i$ is in the incoming edge list of $k$) and from the vertex $k$ to the vertex $j$ ($j$ is in the outgoing edge list of $k$) both edges will have non-$\infty$ values. Thus, there will be no useless relaxation attempt.
\end{pf}

The new algorithm corresponding to the improved form of the Floyd-Warshall algorithm is given in Algorithm 2 \footnote{We also provide very simple C implementation in the Appendix.}.

\begin{algorithm}
  \begin{algorithmic}[1]
  \State $N$: number of vertices
  \State $A[1..N,1..N]$: adjacency matrix ($d_{ij}^0$)
\Procedure {Improved\_FW} {$N, A[1..N,1..N]$}
  \For {$i{\bf :} 1 \to N$} 
  \For {$j{\bf :} 1 \to N$} 
    \If{($i \neq j$) $\wedge$ ($A[i,j]\neq \infty$)} 
      \State {$out[i] \gets out[i] \bigcup \{j\}$}
      \State {$in[j] \gets in[j] \bigcup \{i\}$}
    \EndIf
    \EndFor
    \EndFor
  \For {$k{\bf :} 1 \to N$} 
  \For {{\bf each} $i \in in[k]$} 
  \For {{\bf each} $j \in out[k] $} 
    \If{$A[i,j]>A[i,k]+A[k,j]$} 
      \If {$A[i,j]= \infty$}
      \State {$out[i] \gets out[i] \bigcup \{j\}$}
      \State {$in[j] \gets in[j] \bigcup \{i\}$}
      \EndIf      
      \State {$A[i,j] \gets A[i,k]+A[k,j]$}
    \EndIf
    \EndFor
    \EndFor
  \EndFor
\EndProcedure
 \caption{Improved\_FW}\label{New_FW}
\end{algorithmic}
\end{algorithm}

Note that in the implementation of the union operations ($\bigcup$) since the added vertex is not in the current list yet, it will be simple append operation to the end of the list, and thus, it will take a constant time. The preliminary loops (lines 4-10) generate $in^0$ and $out^0$. The main part (lines from 12 to 24) corresponds to the new recurrence relations. The main differences from the original Floyd-Warshall algorithm are as follows:
\begin{itemize}
    \item The second and the third loops  explore only the incoming and the outgoing edges of the vertex k, rather than exploring the whole matrix. Thus, for sparse graphs, these loops are going to iterate much less than the original Floyd-Warshall algorithm.
    \item Each time a relaxation is discovered between a pair of vertices $i$ and $j$, before it is applied, the incoming and the outgoing lists of the vertices $i$ and $j$ are extended with each other respectively. So that these lists corresponds to the incoming and outgoing edge lists for the matrix $D^k$. 
\end{itemize}

Notice that, as the iterations are applied $in$ and $out$ lists get larger. Therefore, even in the improved algorithm, sometimes the second and the third loops may iterate as many times as closer to the value $N$. A simple way to further improve the performance of this algorithm will be selecting the next vertex $k$ at each iteration by considering the number of its incoming and outgoing edges, rather than processing vertices in the order of 1 to $N$. Since the behaviour of the two inner loops depend on these values, we can choose the next vertex $k'$ having the smallest value for the multiplication of the number of the incoming edges and the number of the outgoing edges of that vertex.  This heuristic improves the performance of two inner loops, with the trade-off on determining the vertex with the  minimum value on each iteration of the outer loop. So, the line 12 will be replaced as follows:

\begin{algorithm}
  \begin{algorithmic}[1]
\State{...}
\State {$all_k \gets \{1, 2, ..., N\}$}
\While {$all_k \neq \emptyset$}
\State {$k' \gets min\{in[k']\times out[k']: \forall k' \in all_k\}$}
  \State {$all_k \gets all_k - k'$} 
\State{...}
\EndWhile
 \caption{Further\_Improved\_FW}\label{Heuristic_New_FW}
\end{algorithmic}
\end{algorithm}

In order to illustrate the behaviour of the proposed algorithm, which is also improved with the above heuristic, we will use the following example adjacency matrix representing a graph with 5 vertices.\\

\begin{tabular}{lllll}
0 & 6 & $\infty$ & 5 & $\infty$ \\
2 & 0 & 3 & -1 & 2 \\
-2 & $\infty$ & 0 & 2 & $\infty$\\
-1 & 1 & 2 & 0 & -1 \\
1 & $\infty$ & $\infty$ & $\infty$ & 0 \\
\end{tabular} \\

Using the above heuristic, vertex 5 will be chosen as the first vertex since it has only 2 incoming edges and 1 outgoing edge, and their multiplication is the minimum among all vertices. Thus, the total number of the relaxation attempts corresponding to the inner loops of the algorithm will also be $2 \times 1 = 2$ for this iteration. Since neither of these two attempts will relax any value, the matrix  will still be the same as the original adjacency matrix after this iteration. 

In the next iteration of the outer loop, vertex 3 will be selected, with 2 incoming and 2 outgoing edges. Therefore, in this iteration, there will be 4 relaxation attempts. Only one of them will be successful, and it will generate a change in the matrix using the relaxation $d_{21}=d_{23}+d_{31}$. The original value of $d_{21}=2$ will be relaxed as  $d_{21}=3+(-2)=1$. The next vertex to be selected will be vertex 1, with 4 incoming edges and 2 outgoing edges. After 8 relaxation attempts in this iteration, 3 previously $\infty$ values of the matrix will become non-$\infty$ values. They are $d_{32}=4,~d_{52}=7,~d_{54}=6$. The new matrix will be as follows:\\

\begin{tabular}{lllll}
0 & 6 & $\infty$ & 5 & $\infty$ \\
1 & 0 & 3 & -1 & 2 \\
-2 & 4 & 0 & 2 & $\infty$\\
-1 & 1 & 2 & 0 & -1 \\
1 & 7 & $\infty$ & 6 & 0 \\
\end{tabular} \\

As a result of these operations, i.e., by converting some $\infty$ values to non-$\infty$ ones,  incoming and outgoing lists of some vertices will also be affected. Namely, incoming list of vertex 2 will get vertex 3 and 5, in addition to previously having only vertex 4, and, incoming list of vertex 4 will also get vertex 5, in addition to previously  containing 1, 2, and 3. Furthermore, outgoing lists of vertex 3 will get vertex 2, in addition to 1 and 4, and finally the outgoing list of vertex 5 will get vertices 2 and 5, in addition to vertex 1. 

Only two vertices are left for the following 2 iterations, which are vertices 2 and 4. Both of them have 4 incoming and 4 outgoing edges. So, they can be processed in any order generating 16+16=32 relaxation attempts, with 4+8=12 successful relaxations. The final matrix is shown below: \\

\begin{tabular}{lllll}
0 & 6 & 7 & 5 & 4 \\
-2 & 0 & 1 & -1 & -2 \\
-2 & 3 & 0 & 2 & 1\\
-1 & 1 & 2 & 0 & -1 \\
1 & 7 & 8 & 6 & 0 \\
\end{tabular}\\

The total number of relaxation attempts will be:\\

2+4+8+16+16=46 \\

The original Floyd Warshall algorithm would have 125 attempts. As it can be seen even from this small and dense matrix example, we can observe significant amount of reduction in the number of useless relaxation attempts. 

In terms of the complexity, for the worst case, there is no improvement. However, all our experiments show that except very dense graphs, the performance of our algorithm is always significantly better than the Floyd-Warshall algorithm. Due to additional overheads, only for very dense graphs our algorithms performs slightly worse than the Floyd-Warshall algorithm. Also, for the initialization of the incoming lists and the outgoing lists, there is an additional preprocessing cost of $N^2$ in our algorithm. Furthermore, selecting the best next vertex to be processed require linear time, and, since it is done at every iteration of the outer loop, there is also another additional cost of $N^2$. 

In terms of the memory requirement, due to additional incoming and outgoing edge lists, our algorithm requires extra memory, which may grow through the iterations. In the worst case, if all vertices can reach all other vertices (whenever this case is achieved at some iteration), the memory requirement will become $3\times N^2$. However, for the incoming and outgoing edge lists since only vertex identifiers are stored (not the wights as in the adjacency matrix), the actual memory requirement might still be much less than 3 times the size of the adjacency matrix. In general, the memory requirement depends on the final matrix structure. If there are $M$ non-$\infty$ values (i.e., edges) in the final matrix, which is expected to be much smaller than $N^2$ for sparse graphs, the maximum memory requirement will be as much as $N^2+2M$ ( $M$ for each one of the incoming and outgoing edge lists).

\section{Experiments}

We have tested our proposed algorithm for some inputs, and, compared its performance against the Johnson's algorithm for graphs with negative weights (assuming no negative cycle), and, against the Dijkstra's algorithm (running it $N$ times using every vertex as a source) as well for graphs with only non-negative weights. Our results show that, especially for sparse graphs our algorithm has superior performance even though these algorithms technically have better worst case complexities. 

We had used standard implementations of these algorithms. Namely, for the Dijkstra's algorithm, we have used the adjacency list representation for storing and processing the graph. Also, we have used the binary heap for storing the current distance values of the vertices, and thus, both selecting the next vertex with minimum distance value and relaxing the values of the vertices are done in logarithmic time. As a result, its complexity is $O(NlgN+MlgN)$, where $M$ is the number of the edges in the graph. Thus, totally, it costs $O(N(NlgN+MlgN))$ for determining the all pairs shortest paths using Dijkstra's algorithm. For implementing the Johnson's algorithm, the same Dijskstra's algorithm is used, and, the Bellman-Ford \cite{Bellman1958}-\cite{Ford1959} algorithm has been implemented with adjacancey list representation as well. Thus, its complexity is $O(NM+N(NlgN+MlgN))$ (plus the additional small reweighting costs of the vertices with $N$). Basically we have utilized the descriptions from \cite{CLRS2009} for the implementation of these algorithms.

\begin{table*}[b]
\caption{Execution times in terms of the percentages of the Floyd-Warshall algorithm}\label{tbl1}
\begin{tabular*}{\tblwidth}{@{} LLLLLLLLLL@{} }
\toprule
Number of Edges & N/2 & N & 2N & 4N & lgN.N & 2lgN.N & (N/lgN).N & (N/2).N\\
\midrule
Dijkstra's & 2\% & 2.4\% & 5.2\% & 5.5\% & 5.6\% & 5.9\% & 18.5\% & 72.8\% \\
Johnson's & 2.7\% &  3.2\% & 6\% & 6.6\% & 7.6\% & 10.1\% & 31.8\% & 136.6\% \\
Ours & 0.4\% & 0.7\% & 7.6\% & 38.2\% &97.9\% & 151.9\% & 196.3\% &  207.4\%\\
\bottomrule
\end{tabular*}
\end{table*}

We have compared our proposed algorithm against the Floyd-Warshall, Dijkstra's (run for each vertex as a source) and Johnson's algorithms as follows:
\begin{itemize}
    \item We have determined the execution times of Dijkstra's, Johnson's and our algorithm in terms of the percentage of the Floyd-Warshall algorithm.
    \item We have created random graphs and used the same graphs for all algorithms. Thus, for all tests we have used graphs with non-negative weights only since the Dijkstra's algorithm's constraint.
    \item For each case, we have generated 10 random graphs and determined the average of these 10 inputs.
    \item We have used sparse and dense graphs with the number of edges, $M$, chosen as follows:\\
    $\{N/2, N,~ 2N,~ 4N,~ lgN.N,~ 2lgN.N, \\ 4lgN.N,~ (N/lgN).N,~ (N/2).N \}$
    \end{itemize}

The results corresponding to graph size as $N=1024$ are shown in the Table 1. As it can be seen from this table, for small graphs (with less than 2N edges) our proposed algorithm performs significantly better than Johnson's algorithm. Even for graphs with only positive weighted edges, it performs better than Dijkstra's algorithm. As graphs gets denser, as it is expected, the alternatives start to perform better. 

We have also compared three algorithms for larger graphs ($N=2048$ and beyond). We have excluded the original Floyd-Warshall algorithm for larger graphs, since it was taking too much time. The results of these experiments show that our algorithm benefits more than Dijkstra and Johnson's algorithms for similar sparse graph structures due to the overhead that comes from the complex data structures of Dijkstra and Johnson's algorithms..   

\section{Conclusions}

In this paper we proposed a simple improvement to the Floyd-Warshall algorithm. We have shown that with this improvement all unnecessary relaxation attempts of the Floyd-Warshall algorithm can be avoided. In addition, using a heuristic for selecting the next vertex to apply the reduction operation through it, also further reduced the number of reduction attempts. Through experiments, we  show the effectiveness of this approach, especially for sparse and large graphs. 

\newpage
\appendix
\section{Implementation of the Improved Floyd-Warshall Algorithm in C}
\begin{verbatim}
/*	Ismail Hakki Toroslu
	September 1, 2021
	Improved Floyd-Warshall Algorithm

	Two lists, inlist and outlist, can be implemented 
	more efficiently as linked lists.
	For simplicity simple array implementation is used.
*/
#include <stdio.h>
#include <stdlib.h>
#define MAXN 1025    // Maximum number of nodes
#define INF 9999     // Used as inifnity
 
int A[MAXN][MAXN];   // Adjaceny matrix

void new_fw(int N)
{
    int inc[MAXN],outc[MAXN];   // counts of in/out edges
    // List of vertices incoming/outgoing to/from 
    int inlist[MAXN][MAXN], outlist[MAXN][MAXN];   
    int i,j,k,kk;
    int select_k[MAXN], mininxout, mink;  // choose the "best" k

    for (i=0;i<N;i++)
        inc[i]=0,outc[i]=0,select_k[i]=0;

    // Generate initial inlist and outlist for each vertex
    for (i=0;i<N;i++)
       for (j=0;j<N;j++)
       {
         if ((A[i][j]!=0) && (A[i][j]<INF))
            inc[j]++, outc[i]++, 
            inlist[j][inc[j]-1]=i, outlist[i][outc[i]-1]=j;
       }

    // The outer loop
    for (kk=0;kk<N;kk++)
    {
        // choose the "best" k
        mink=-1;
        mininxout=2*N*N;
        for (k=0;k<N;k++)
        {
           if ((select_k[k]==0) &&(inc[k]*outc[k]<mininxout))
           {
              mink=k;
              mininxout=inc[k]*outc[k];
           }
        }
        k=mink;         // "best" k
        select_k[k]=1;  // remove selected vertex
        
        // explore only useful relaxation attempts
        for (i=0;i<inc[k];i++)
            for (j=0;j<outc[k];j++)
            {
                if ((A[inlist[k][i]][k]+A[k][outlist[k][j]])<
                      A[inlist[k][i]][outlist[k][j]])
                {
                   if (A[inlist[k][i]][outlist[k][j]]==INF) 
                   {
                      outc[inlist[k][i]]++;
                      outlist[inlist[k][i]][outc[inlist[k][i]]-1]=
                         outlist[k][j];
                      inc[outlist[k][j]]++;
                      inlist[outlist[k][j]][inc[outlist[k][j]]-1]=
                         inlist[k][i];
                   }
                   A[inlist[k][i]][outlist[k][j]]=
                      A[inlist[k][i]][k]+A[k][outlist[k][j]];
               }
           }
     }
     
     printf("Improved FW APSP\n");
     for (i=0;i<N;i++,printf("\n"))
        for (j=0;j<N;j++)
            if (A[i][j] != INF) printf("%d ",A[i][j]);
            else printf("X ");  // print "X" instead of infinity
} 

int main()
{
    int N, i, j;
    
    // input N and adjacency matrix (infinity is 9999)
    scanf("%d",&N);
    for(i=0;i<N;i++)
        for(j=0;j<N;j++)
           scanf("%d",&A[i][j]);

    new_fw(N);
    return 0;
}
\end{verbatim}
\printcredits

\bibliographystyle{cas-model2-names}

\bibliography{cas-refs}


\bio{}

\endbio

\end{document}